%% file: main.tex
\documentclass[12pt,a4paper]{article}

\usepackage{xcolor}
 
\usepackage{ifthen} 
\newboolean{pdflatex}
\setboolean{pdflatex}{true} 

\newboolean{articletitles}
\setboolean{articletitles}{true} 

\newboolean{uprightparticles}
\setboolean{uprightparticles}{false} 

\newboolean{inbibliography}
\setboolean{inbibliography}{false} 
  
\usepackage{microtype}
\usepackage{lineno}  
\usepackage{xspace} 
\usepackage{caption} 

\usepackage{graphicx}  
\usepackage{color}
\usepackage{colortbl}
\graphicspath{{./figs/}} 

\usepackage{amsmath} 
\usepackage{amssymb}
\usepackage{amsfonts}
\usepackage{upgreek} 

\newcommand*\patchAmsMathEnvironmentForLineno[1]{%
\expandafter\let\csname old#1\expandafter\endcsname\csname #1\endcsname
\expandafter\let\csname oldend#1\expandafter\endcsname\csname
end#1\endcsname
 \renewenvironment{#1}%
   {\linenomath\csname old#1\endcsname}%
   {\csname oldend#1\endcsname\endlinenomath}%
}
\newcommand*\patchBothAmsMathEnvironmentsForLineno[1]{%
  \patchAmsMathEnvironmentForLineno{#1}%
  \patchAmsMathEnvironmentForLineno{#1*}%
}
\AtBeginDocument{%
\patchBothAmsMathEnvironmentsForLineno{equation}%
\patchBothAmsMathEnvironmentsForLineno{align}%
\patchBothAmsMathEnvironmentsForLineno{flalign}%
\patchBothAmsMathEnvironmentsForLineno{alignat}%
\patchBothAmsMathEnvironmentsForLineno{gather}%
\patchBothAmsMathEnvironmentsForLineno{multline}%
\patchBothAmsMathEnvironmentsForLineno{eqnarray}%
}

\usepackage{hyperref}    
\usepackage[all]{hypcap} 

\input{lhcb-symbols-def} 

\usepackage{cite} 
\usepackage{mciteplus}

\usepackage{longtable} 

\usepackage{graphicx}
\usepackage{subcaption}

\newcommand{\BRof}[1]{\ensuremath{{\cal B}(#1)}\xspace}
\newcommand{\Dtoemu}{\ensuremath{D^0 \to e^\pm \mu^\mp}\xspace}
\newcommand{\DtoKpi}{\ensuremath{D^0 \to \Km \pip}\xspace}
\newcommand{\Dtopipi}{\ensuremath{D^0 \to \pip \pim}\xspace}
\newcommand{\DstarToDpi}{\ensuremath{\Dstarp \to \Dz \pip}\xspace}
\newcommand{\DstarToDemupi}{\ensuremath{\Dstarp \to \Dz (\epm \mump) \pip}\xspace}
\newcommand{\DstarToDKpipi}{\ensuremath{\Dstarp \to \Dz (\Km \pip) \pip}\xspace}
 
\def\mump        {{\ensuremath{\Pmu^\mp}}\xspace} 
\def\CLs           {CL$_{\text S}$}
\newcommand{\Btoemu}{\ensuremath{B^0 \to e^\pm \mu^\mp}\xspace}
\newcommand{\Bstoemu}{\ensuremath{B_s^0 \to e^\pm \mu^\mp}\xspace}
\begin{document}

\renewcommand{\thefootnote}{\fnsymbol{footnote}}
\setcounter{footnote}{1}

\begin{titlepage}
\pagenumbering{roman}

\vspace*{-1.5cm}
\centerline{\large EUROPEAN ORGANIZATION FOR NUCLEAR RESEARCH (CERN)}
\vspace*{1.5cm}
\noindent
\begin{tabular*}{\linewidth}{lc@{\extracolsep{\fill}}r@{\extracolsep{0pt}}}
\ifthenelse{\boolean{pdflatex}}
{\vspace*{-2.7cm}\mbox{\!\!\!\includegraphics[width=.14\textwidth]{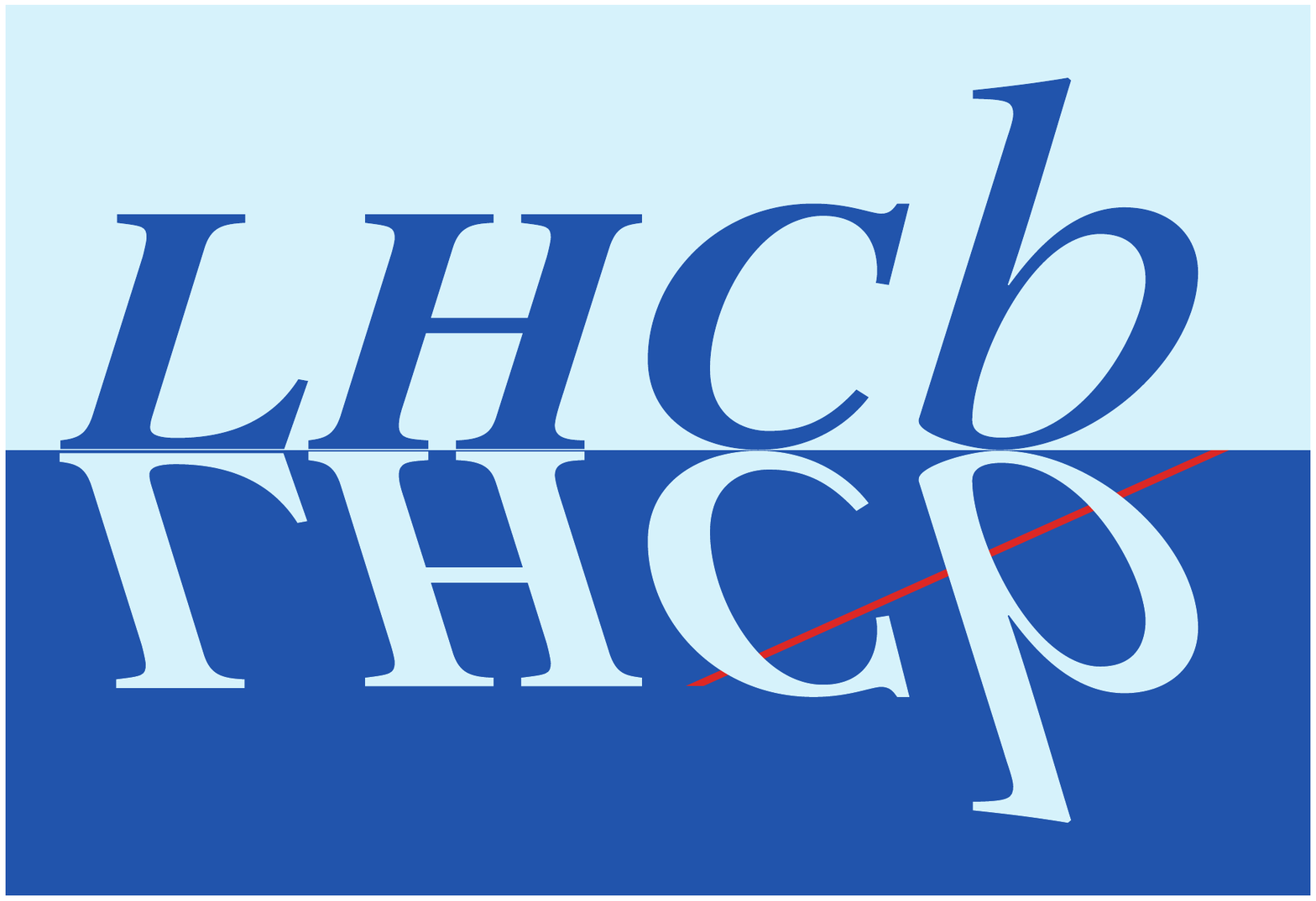}} & &}
{\vspace*{-1.2cm}\mbox{\!\!\!\includegraphics[width=.12\textwidth]{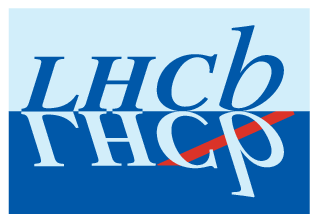}} & &}
\\
 & & CERN-PH-EP-2015-306 \\
 & & LHCb-PAPER-2015-048 \\ 
 & & 26 January 2016 \\
\end{tabular*}

\vspace*{3.0cm}

{\bf\boldmath\huge
\begin{center}
  Search for the lepton-flavour violating decay \Dtoemu
\end{center}
}

\vspace*{1.5cm}

\begin{center}
The LHCb collaboration\footnote{Authors are listed at the end of this paper.}
\end{center}

\vspace{\fill}

\begin{abstract}
  \noindent
A search for the lepton-flavour violating decay \Dtoemu is
made with a dataset corresponding to an integrated luminosity
of \mbox{3.0 \invfb} of proton-proton collisions at centre-of-mass energies of $7$\,\tev and $8$\,\tev, collected by the LHCb experiment.
Candidate $\Dz$ mesons are selected using the decay $\Dstarp \to \Dz \pip$ and the \Dtoemu branching fraction
is measured using the decay mode $\Dz \to \Km \pip$ as a normalisation channel.
No significant excess of \Dtoemu candidates over the expected background is seen, and a
limit is set on the branching fraction, $\BRof \Dtoemu  < 1.3  \times 10^{-8}$, at 90\% confidence level.
This is an order of magnitude lower than the previous limit and it further constrains the
parameter space in some leptoquark models and in supersymmetric models with R-parity violation.
  
\end{abstract}

\vspace*{2.0cm}

\begin{center}
  Published in Phys.~Lett.~B754~(2016)~167   
\end{center}

\vspace{\fill}

{\footnotesize 
\centerline{\copyright~CERN on behalf of the \lhcb collaboration, licence \href{http://creativecommons.org/licenses/by/4.0/}{CC-BY-4.0}.}}
\vspace*{2mm}

\end{titlepage}


\newpage
\setcounter{page}{2}
\mbox{~}


\renewcommand{\thefootnote}{\arabic{footnote}}
\setcounter{footnote}{0}



\pagestyle{plain} 
\setcounter{page}{1}
\pagenumbering{arabic}


%
 
\section{Introduction}
\label{sec:Introduction}

Searches for decays that are forbidden in the Standard Model (SM) probe potential contributions from new processes and 
particles at mass scales beyond the reach of direct searches. The  decay \mbox{\Dtoemu}  
is an example of a forbidden decay, in which lepton flavour is not conserved.\footnote{The inclusion of charge-conjugate 
processes is implied.} The contributions to this process from neutrino oscillations would give a rate that is
well below the reach of any currently feasible experiment.
However, the decay is predicted to occur in several other models that extend the SM, with rates varying by 
up to eight orders of magnitude. 

In Ref.~\cite{Burdman:2001tf} three extensions to the SM are considered: in a
minimal supersymmetric (SUSY) SM with R-parity violation (RPV) the branching fraction $\BRof \Dtoemu$ 
could be as large as $\cal{O}$$(10^{-6})$; in a theory with multiple Higgs doublets it would be less than about $7\times 10^{-10}$;
and in the SM extended with extra fermions the branching fraction would be less than $\cal{O}$$(10^{-14})$. In Ref.~\cite{Tahir:2014ura}
an RPV SUSY model is considered in which limits on products of couplings are obtained from the experimental
upper limit on the branching fraction $\BRof {\Dsp \to \Kp \epm \mump}$; from these limits,  
$\BRof \Dtoemu$ could be as large as $3 \times 10^{-8}$. A similar study of constraints on coupling constants in 
RPV SUSY~\cite{Wang:2014dba}, obtained from limits on the branching fraction $\BRof {\Dp \to \pip \epm \mump}$,
showed that $\BRof \Dtoemu$ could reach $10^{-7}$.
LHCb has previously set limits~\cite{LHCb-PAPER-2013-030} on branching fractions for the $B$ meson decays \Btoemu and \Bstoemu, using them to put 
lower limits on the masses of Pati-Salam leptoquarks~\cite{Pati:1974yy}. As is shown in Ref.~\cite{Valencia:1994cj}, lepton-flavour violating
charm decays are relatively insensitive to the presence of such leptoquarks. However, in a recent paper~\cite{deBoer:2015boa} it is shown that in
other leptoquark scenarios $\BRof \Dtoemu$ could be as large as $4 \times 10^{-8}$.

The first experimental limit on $\BRof \Dtoemu$ was from Mark II~\cite{10.1103/PhysRevD.35.2914}, and more 
recent results have come from
E791~\cite{10.1016/S0370-2693(99)00902-8} and BaBar~\cite{Lees:2012jt}. 
The most stringent limit is from Belle~\cite{Petric:2010yt}, $\BRof \Dtoemu <2.6\times{}10^{-7}$ at $90\%$ confidence 
level (CL). An improved limit, below $\cal{O}$$(10^{-7})$, would provide tighter constraints on coupling constants in RPV SUSY models~\cite{Burdman:2001tf, Tahir:2014ura, Wang:2014dba}, while a limit below $4 \times 10^{-8}$ would also constrain the parameter space in some leptoquark models~\cite{deBoer:2015boa}.

This Letter presents a search for the decay \Dtoemu using $pp$ collision data corresponding to integrated luminosities
of \mbox{1.0\,\invfb} at a centre-of-mass energy of $7$\,\tev and \mbox{2.0\,\invfb} at $8$\,\tev,
collected by the LHCb experiment in 2011 and 2012, respectively. 
In the analysis, signal candidates are selected using the decay \DstarToDpi and the measurements are normalized 
using the well-measured channel \DtoKpi, which has the same topology as the signal.
A multivariate analysis based on a boosted
decision tree algorithm (BDT) is used to help separate signal and background. 
The mass spectrum in the signal 
region, defined as $1815 - 1915$\,\mevcc, is not examined until all analysis choices are finalized. 

\section{Detector and simulation}
\label{sec:Detector}

The \lhcb detector~\cite{Alves:2008zz,LHCb-DP-2014-002} is a single-arm forward
spectrometer covering the \mbox{pseudorapidity} range $2<\eta <5$,
designed for the study of particles containing \bquark or \cquark
quarks. The detector includes a high-precision tracking system
consisting of a silicon-strip vertex detector surrounding the $pp$
interaction region,
a large-area silicon-strip detector located
upstream of a dipole magnet with a bending power of about
$4{\rm\,Tm}$, and three stations of silicon-strip detectors and straw
drift tubes
placed downstream of the magnet.
The tracking system provides a measurement of momentum, \ptot, of charged particles with
a relative uncertainty that varies from $0.5\%$ at low momentum to $1.0\%$ at 200\gevc.
The minimum distance of a track to a primary vertex (PV), the impact parameter, 
is measured with a resolution of $(15+29/\pt)\mum$,
where \pt is the component of the momentum transverse to the beam, in\,\gevc.
Different types of charged hadrons are distinguished using information
from two ring-imaging Cherenkov detectors.
Photons, electrons and hadrons are identified by a calorimeter system consisting of
scintillating-pad and preshower detectors, an electromagnetic
calorimeter and a hadronic calorimeter. Muons are identified by a
system composed of alternating layers of iron and multiwire
proportional chambers.

The online event selection is performed by a trigger~\cite{LHCb-DP-2012-004}, 
which consists of a hardware stage, based on information from the calorimeter and muon
systems, followed by a software stage in which all charged particles
with $\pt>500\,(300)\mevc$ are reconstructed for 2011\,(2012) data.  At the hardware trigger stage, events are required to have a muon 
with high \pt, or a hadron, photon or electron with high transverse energy in the calorimeters. The software trigger requires a two-, three-
or four-track secondary vertex with a significant displacement from the primary $pp$ interaction vertices.
At least one charged particle must have a transverse momentum $\pt > 1.7\gevc$ and be
inconsistent with originating from a PV. A multivariate algorithm~\cite{BBDT} is used for
the identification of secondary vertices consistent with the decay
of a \bquark or \cquark hadron. 
 
In the simulation, $pp$ collisions are generated using
\pythia~\cite{Sjostrand:2007gs,*Sjostrand:2006za} 
with a specific \lhcb
configuration~\cite{LHCb-PROC-2010-056}.  Decays of hadronic particles
are described by \evtgen~\cite{Lange:2001uf}, in which final-state
radiation is generated using \photos~\cite{Golonka:2005pn}. The
interaction of the generated particles with the detector, and its response,
are implemented using the \geant
toolkit~\cite{Allison:2006ve, *Agostinelli:2002hh} as described in
Ref.~\cite{LHCb-PROC-2011-006}. Samples of simulated events are generated for
the signal \Dtoemu channel, for the normalization \DtoKpi channel and for \Dtopipi, which
is an important background channel.
  
\begin{figure}[!t]
\centering
\includegraphics[width=0.7\textwidth]{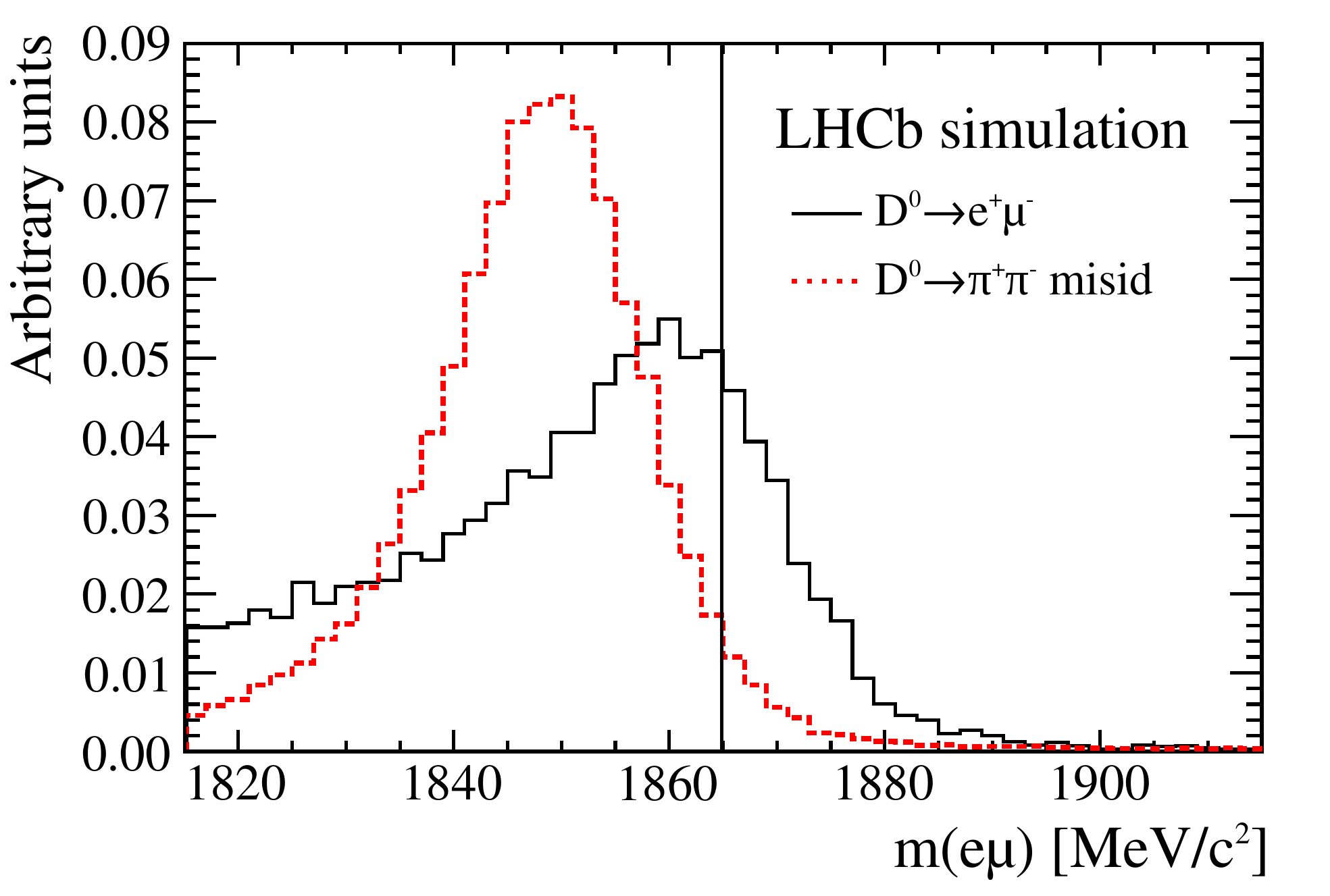}

\caption{Mass spectra from simulation for \Dtoemu decays (solid line) and \Dtopipi decays reconstructed 
as \Dtoemu (dashed line). Each spectrum is normalized to unit area. The vertical line indicates 
the mass of the \Dz meson.}
\label{fig:pipiemu}
\end{figure}
	
\section{Event selection and efficiencies}
\label{sec:Selection}

In the first stage of the offline event selection, the \DstarToDemupi and \DstarToDKpipi candidates that pass the trigger selection are required to have 
a vertex, formed from two good-quality tracks associated with particles of opposite charge, that is well separated from any PV, with the summed 
momentum vector of the two
particles pointing to a PV (the mean number of PVs per beam crossing is $1.6$). The measured momentum of the electron candidates is corrected to 
account for loss of momentum by bremsstrahlung in the 
detector, using the photon energy deposition in the electromagnetic calorimeter~\cite{LHCb-PAPER-2013-005}.
Muon and electron candidates, and pions and kaons from the \DtoKpi candidates, are required to have 
$p > 4$\,\gevc and $p_T > 0.75$\,\gevc and to be positively identified by the particle identification systems. 
The soft pion from the candidate \DstarToDpi decay is required to have $p_T > 110$\,\mevc and to be consistent with coming from the PV. 
A kinematic fit is performed, with the two \Dz decay tracks constrained to a secondary vertex and the soft pion
and \Dz candidates constrained to come from the PV. This fit improves the resolution on the 
mass difference between the reconstructed \Dstarp and \Dz mesons, which is required to be in the range 
$135 - 155$\,\mevcc.
About $2\%$ of events contain more than one \DstarToDpi candidate and in these events one is chosen at random.
After the above selections, $2114$ candidates remain in the signal mass region for \Dtoemu and $330\,359$ for \DtoKpi
(the trigger accept rate for the latter channel is scaled to retain only $1\%$ of candidates). 

An important source of background in the sample of \Dtoemu candidates comes from \Dtopipi decays 
where one pion is misidentified as an electron and the other as a muon. From simulations and calibration
samples in the data~\cite{LHCb-DP-2014-002}, the probability for a \Dtopipi event to be selected in the final sample of candidate signal events is 
found to be $(1.0 \pm 0.6) \times 10^{-8}$ in the $7$\,\tev data and $(1.8 \pm 0.4) \times 10^{-8}$ in the $8$\,\tev data. Figure~\ref{fig:pipiemu} shows a comparison of
the mass spectra, from simulation, for \Dtoemu decays and for \Dtopipi decays reconstructed as \Dtoemu, with each spectrum
normalized to unit area. The low-mass tail for genuine \Dtoemu decays is caused by bremsstrahlung from the electrons; about $15\%$
of the signal lies below $1810\,\mevcc$. 
The misidentified  \Dtopipi decays produce
a peak at a mass about $15$\mevcc below the signal mass. 
Misidentified \DtoKpi decays always have reconstructed mass below the region selected for the analysis,
because of the large mass difference between kaons and electrons or muons; as
a consequence, there is no background from
this source. Other sources of background include the
semileptonic decay modes $\Dz \to \pim \ep \neue$ and $\Dz \to \pim \mup \neum$, with the pion misidentified
as a muon or an electron, respectively. Since, as part
of bremsstrahlung recovery, the energy of unrelated photons may be incorrectly added to the energy
of the electron candidates, these semileptonic backgrounds extend smoothly above the signal region
and are treated as part of the combinatorial background of $e^\pm\mu^\mp$
pairs where the two lepton candidates have different sources.

Trigger, selection and particle identification efficiencies, and misidentification probabilities, are obtained from a combination 
of simulation
and data. Control samples of well-identified electrons, muons, pions and kaons in data are obtained
from \jpsi meson decays into pairs of electrons or muons and from \DstarToDKpipi decays, selected using
different requirements from those used in the current analysis.  
These control samples are binned in pseudorapidity and transverse momentum of the tracks, and in the 
track multiplicity of the event.
The hardware trigger efficiency for signal is evaluated using data, while the efficiency for the software trigger and 
offline selections is evaluated 
using simulation after validation with the data control samples. Where efficiencies are taken from the simulation, 
the samples are 
weighted to take into account
differences between simulation and data, particularly in the distribution of per-event track multiplicities. 

\section{Multivariate classifier}
\label{sec:BDT}

A multivariate classifier based on a BDT~\cite{Breiman} 
with a gradient boost~\cite{Friedman:2001} is used to divide the selected sample into bins of
different signal purity. 
The following variables are used as
inputs to the BDT: the smallest distance of closest approach of the \Dz candidate to any PV;
an isolation variable that depends on how much additional charged particle momentum is in a region
of radius $R \equiv \sqrt{(\Delta \eta)^2+(\Delta \phi)^2} = 1$ around the \Dstarp candidate,
where $\eta$ and $\phi$ are pseudorapidity and azimuthal angle; 
$\chi^2$ of the kinematic fit; and \chisqip, the impact parameter $\chi^2$ with respect to 
the  associated PV, for each of the \Dstarp and \Dz candidates, and for the two \Dz decay tracks. The variable \chisqip is defined 
as the difference in vertex fit $\chi^2$ with and without the particle considered.  
None of
the BDT input variables contains particle identification information. It therefore  
performs equally well for the signal and normalization channels (and for the misidentified
\Dtopipi decays).

The BDT is trained separately for the $7$\,\tev and $8$\,\tev data samples, to exploit the dependence
of some input variables, for example the isolation variable, on the collision energy. The background sample used for the
training comprises selected candidates with invariant mass within $300$\,\mevcc of
the known \Dz mass, but excluding the signal region, $1815 - 1915$\,\mevcc. The training for
signal is done with the simulated \Dtoemu events. One half of each sample is used
for training the BDT, while the other half is used to test for over-training. No evidence for over-training is seen.
Following procedures used in Refs.~\cite{LHCb-PAPER-2013-014, LHCb-PAPER-2014-052}, the BDT output value, which lies 
between $-1$ (most background-like) and $1$ (most signal-like),
is used to separate the data sample into three sub-samples
with ranges chosen to
give optimum separation between the background-only and signal-plus-background
hypotheses.

\section{Fits to mass spectra}
\label{sec:Fits}

In order to determine the number of signal decays, extended maximum likelihood fits are made simultaneously to 
unbinned distributions of $m(\Dz)$ and $\Delta m = m(\Dstarp) - m(\Dz)$ for the \Dtoemu candidates
in each of the three BDT bins for the $7$\,\tev and $8$\,\tev data. Hereinafter, $m(\Dz)$ denotes the mass of the \Dz 
candidate for both signal and normalization channels, and $\Delta m$ denotes the mass difference between the 
\Dstarp and \Dz candidates. In these fits, from which the branching fraction is extracted directly, all systematic 
uncertainties, as discussed in Sect.~\ref{sec:systematics}, are included as Gaussian constraints on the appropriate
parameters.

The \Dtoemu signal probability density functions (PDF) in the three BDT bins are obtained from the simulation. The  
simulated \Dtoemu mass 
spectra are fitted using the sum of two Crystal Ball functions~\cite{Skwarnicki:1986xj} with a common peak value but different widths. 
One of the Crystal Ball functions has a low-mass tail to account for energy loss due to 
bremsstrahlung while the other is modified to have a high-mass tail to accommodate events where a bremsstrahlung photon 
is incorrectly assigned to an electron candidate. The per-event particle multiplicity
affects the amount of 
bremsstrahlung radiation recovered for the electron candidates, and this differs between simulation and data.
Therefore both the simulation and the data are classified in three bins of the variable $N_{\rm SPD}$, 
the number of hits in the scintillating pad detector, which is a measure of the particle multiplicity. The parameters
of the signal PDF are obtained as averages of their values in the three bins of $N_{\rm SPD}$, weighted to account
for data-simulation differences.
The PDF shapes for the peaking background
due to misidentified \Dtopipi decays (see Fig.~\ref{fig:pipiemu}) are obtained in the same way as for \Dtoemu, using 
the same functional form for the signal shapes, and their yields are Gaussian-constrained in the fits. The combinatorial background 
for the \Dz candidate mass is described by a second-order polynomial. 

The signal shapes in the $\Delta m$ distributions for the \Dtoemu and \Dtopipi channels are each parametrised as a sum of three Gaussian functions; 
for \Dtoemu two of the Gaussians functions have the same mean, but the one with the largest width is allowed to have a different mean,
while the three mean values are independent for the \Dtopipi shape. In each case all three Gaussian functions have independent widths.
The combinatorial background in $\Delta m$ is fitted using an empirical function of the form
\begin{equation} \label{eq:background}
  f(\Delta m) = N\left[\left(1-\exp\left(-\frac{\Delta m - {(\Delta m)}_0}{c}\right)\right)\times{}\left(\frac{\Delta m}{{(\Delta m)}_0}\right)^a 
+ b\left(\frac{\Delta m}{{(\Delta m)}_0}-1\right)\right]~,
\end{equation}
where $N$ is a normalization factor, ${(\Delta m)}_0$ is the threshold mass difference, and $a$, $b$ and $c$ are free parameters.
In the fits to the \Dtoemu candidates, the parameter $a$ is fixed to zero. 
A fraction of the \Dtoemu and the misidentified \Dtopipi decays is associated to a random soft pion, and therefore peaks in
$m(\Dz)$ but not in $\Delta m$. This fraction is Gaussian constrained  to the value $23.7 \pm 0.2\%$
found in the fits to the \DtoKpi normalisation channel, discussed below.

\begin{figure}[!t]
\centering
  \begin{subfigure}[b]{0.49\textwidth}
    \includegraphics[width=\textwidth]{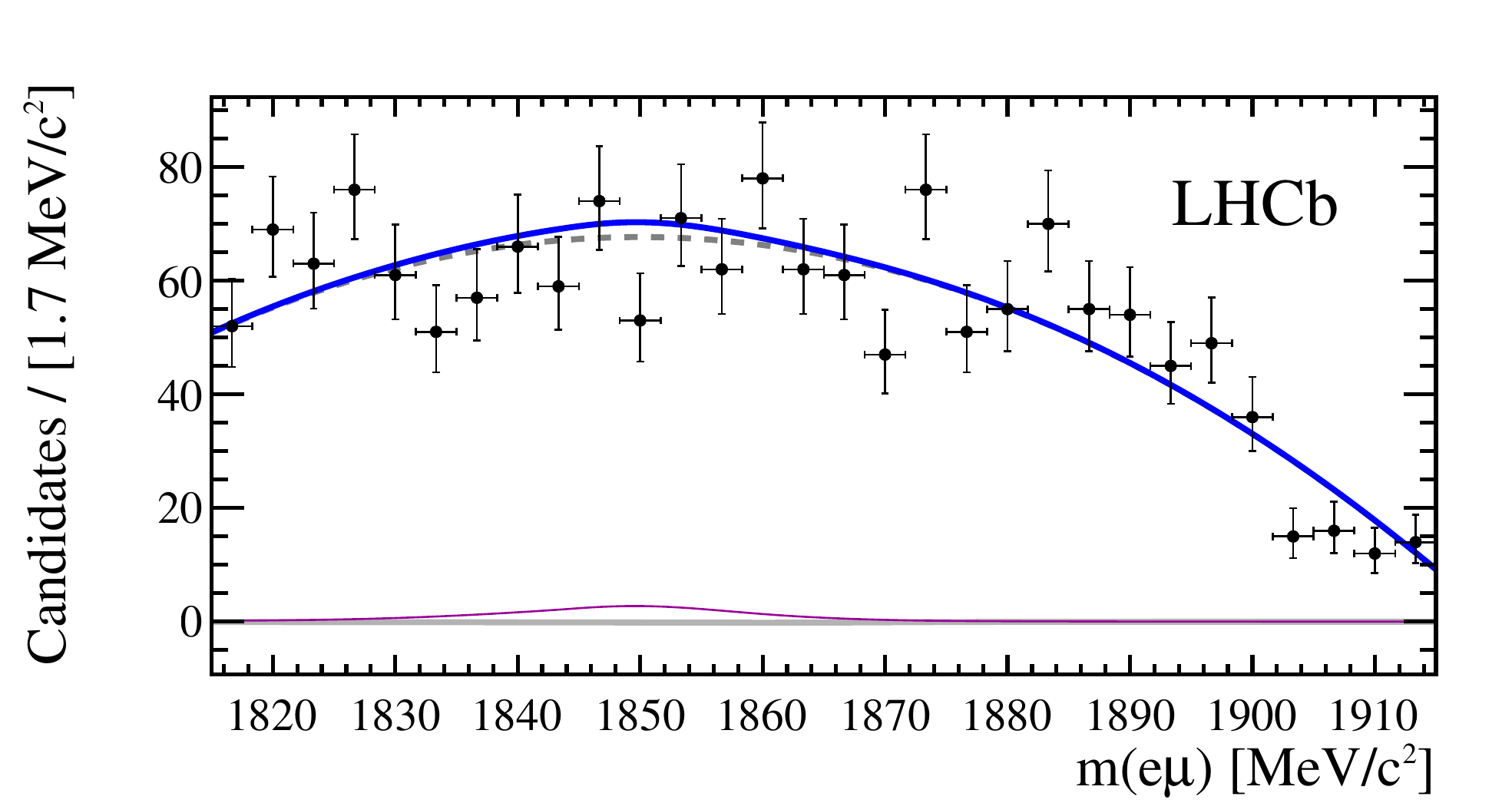}
  \end{subfigure}
\begin{subfigure}[b]{0.49\textwidth}
    \includegraphics[width=\textwidth]{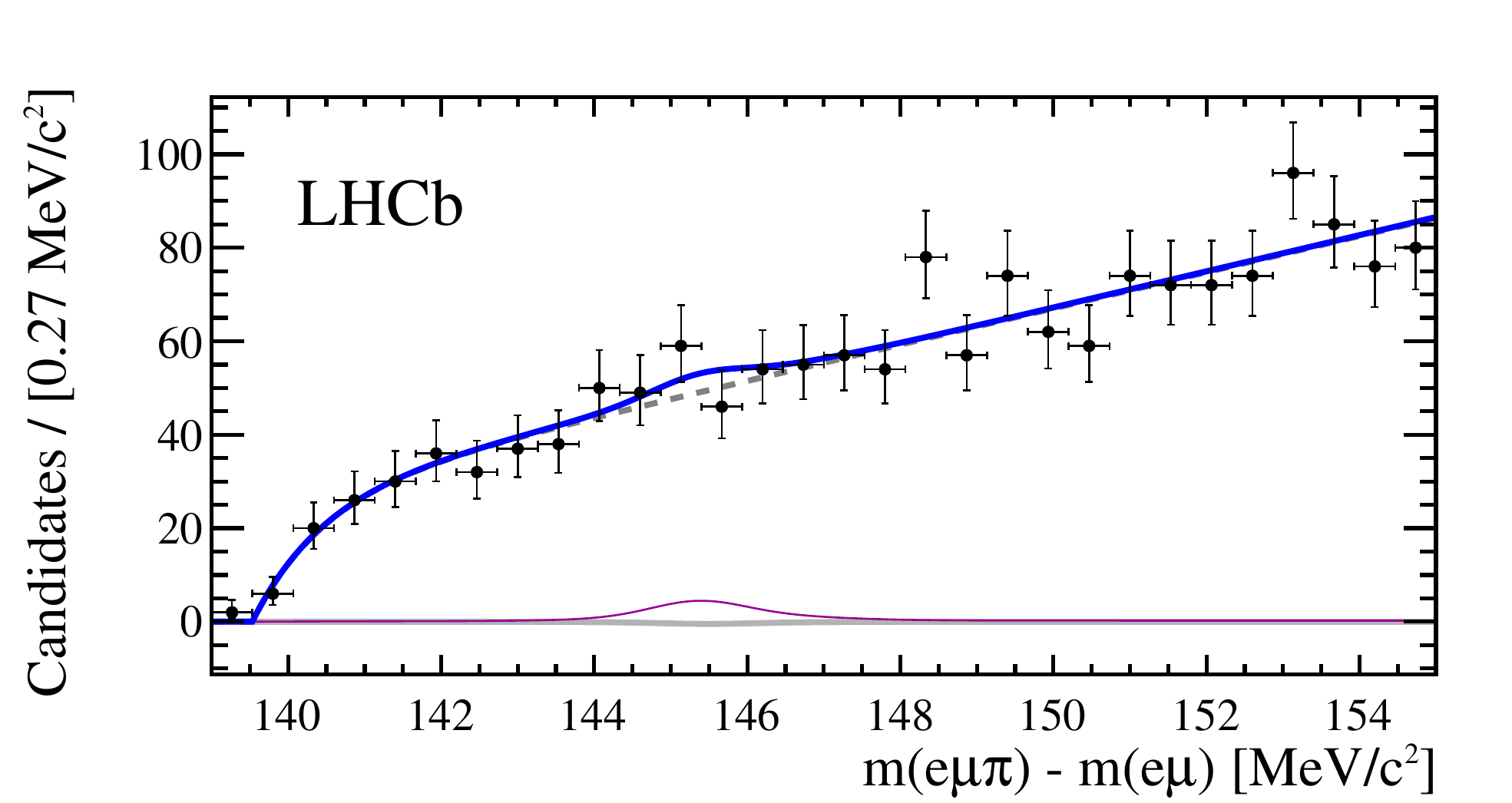}
  \end{subfigure}
\begin{subfigure}[b]{0.49\textwidth}
    \includegraphics[width=\textwidth]{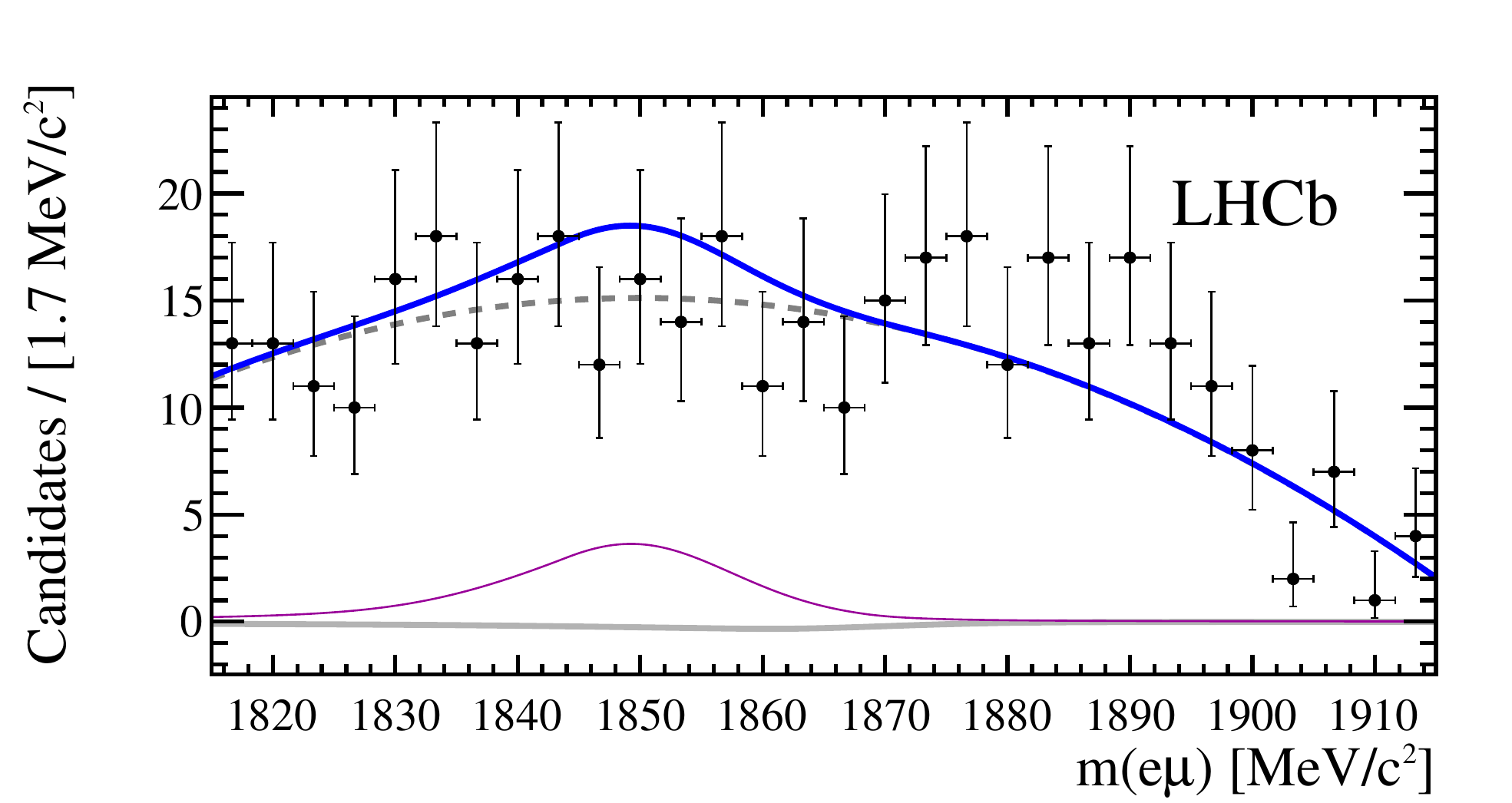}
  \end{subfigure}
\begin{subfigure}[b]{0.49\textwidth}
    \includegraphics[width=\textwidth]{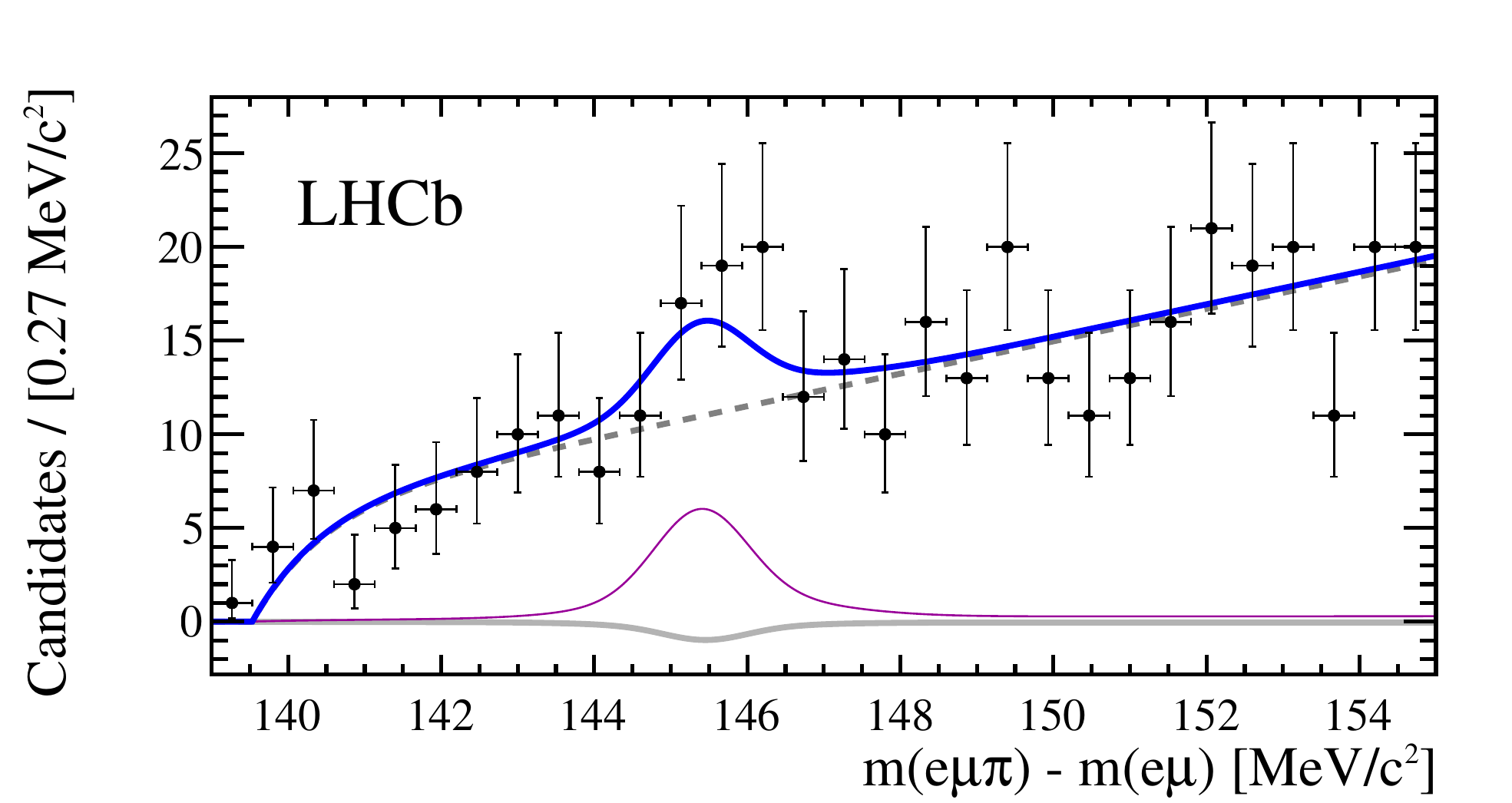}
  \end{subfigure}
\begin{subfigure}[b]{0.49\textwidth}
    \includegraphics[width=\textwidth]{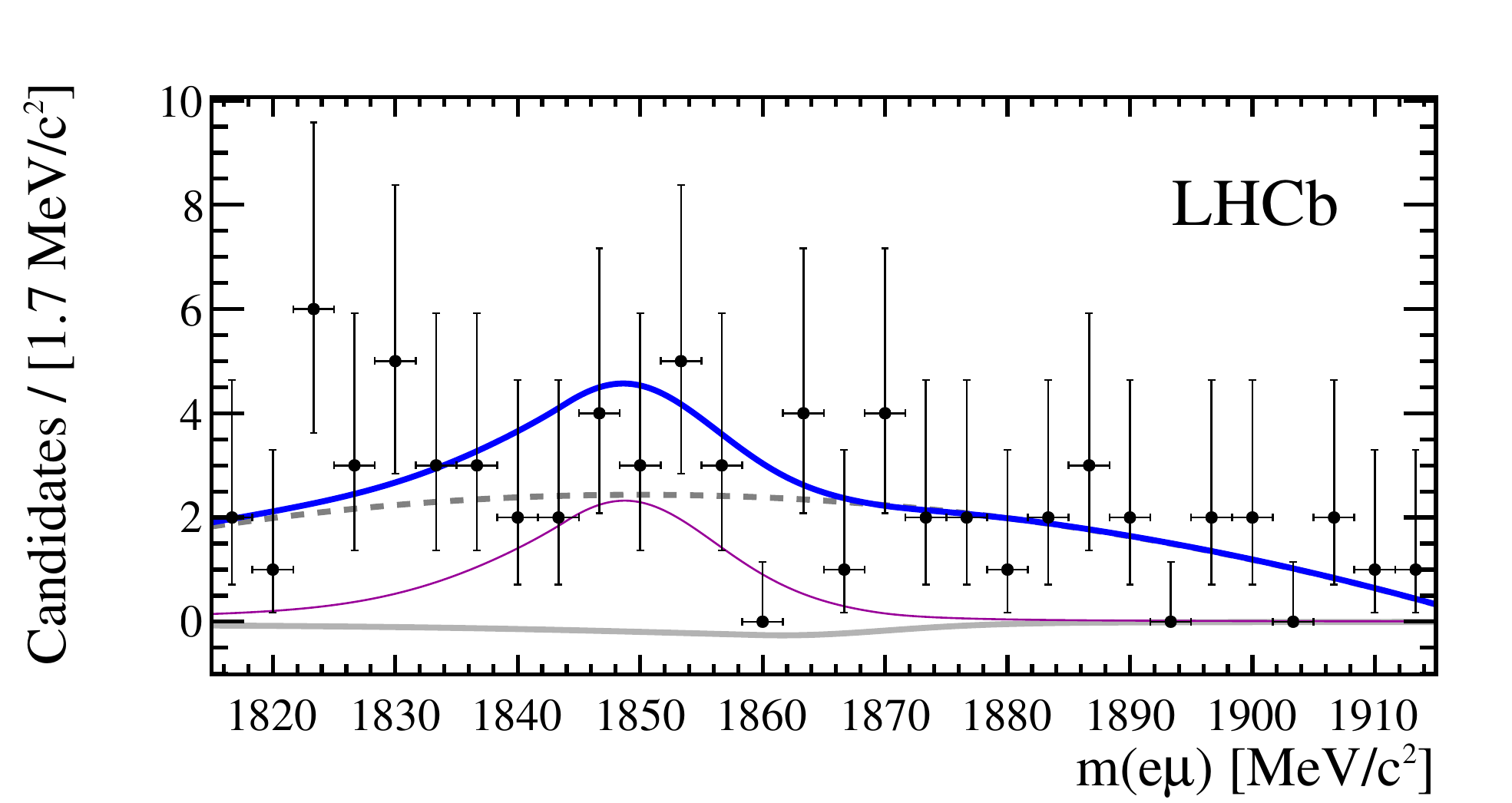}
  \end{subfigure}
\begin{subfigure}[b]{0.49\textwidth}
    \includegraphics[width=\textwidth]{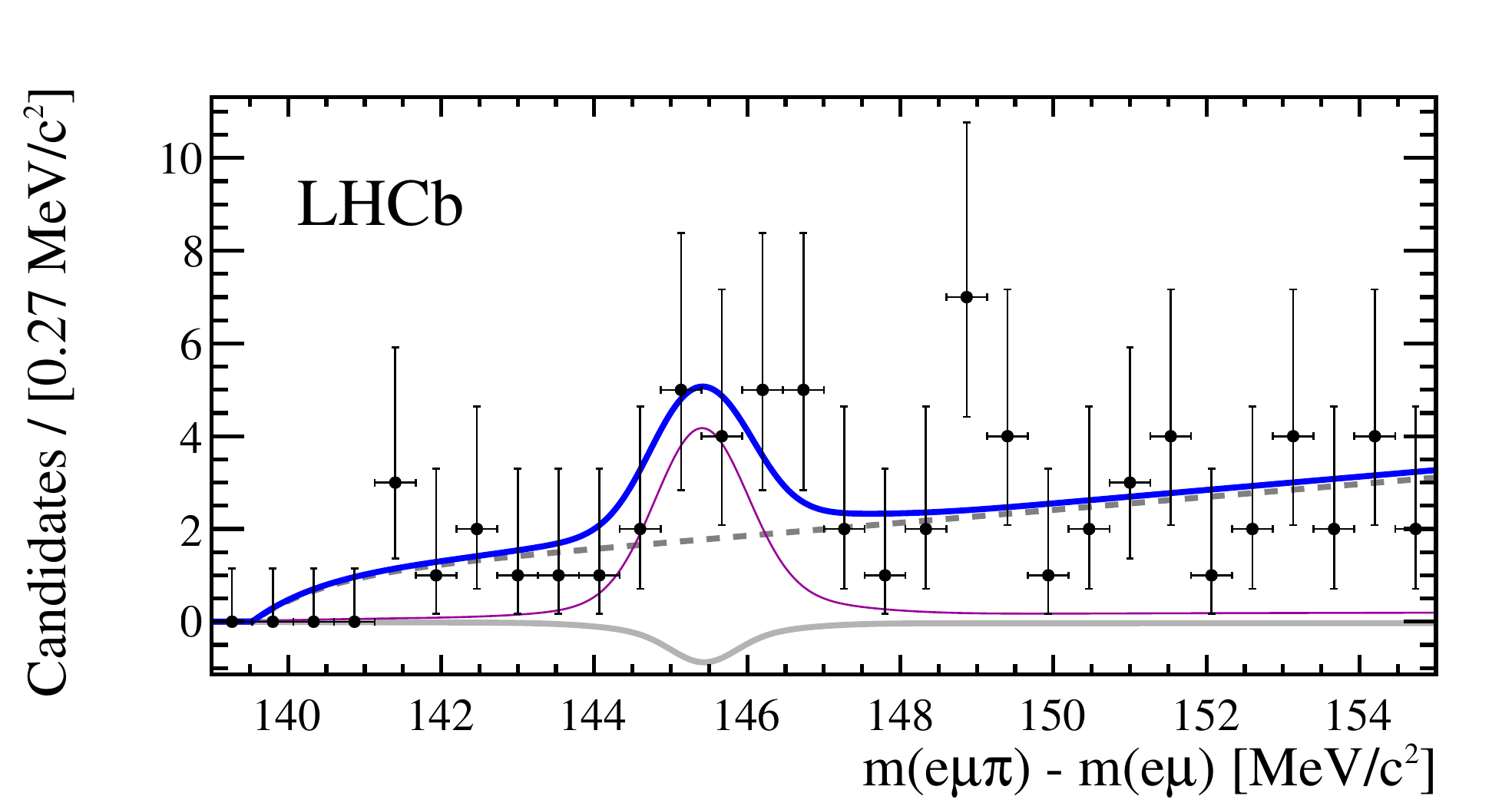}
  \end{subfigure}
    \caption{Distributions of (left) $m(\protect\Dz)$ and (right) $\Delta m$ for \Dtoemu candidates reconstructed in the combined $7$\,\protect\tev and $8$\,\protect\tev data, 
with fit functions overlaid. 
The rows correspond to the three bins of BDT output, with the top row corresponding to the most background-like and the
bottom row to the most signal-like. The solid (blue) lines show the total fit results, while the thick (grey) lines show the total \protect\Dtoemu component,
the thin (purple) lines show the total misidentified \protect\Dtopipi and the dashed (grey) lines indicate the combinatorial background.}
\label{fig:fulldata}
\end{figure}

Figure~\ref{fig:fulldata} shows the fit results for the combined $7$\,\tev and $8$\,\tev dataset, separately for the three bins of 
BDT output. The peaks seen in the $m(\Dz)$ and $\Delta m$ distributions are due to misidentified \Dtopipi decays. No evidence
is seen for any \Dtoemu signal. The fits return a total of $-7 \pm 15$ signal decays.

\begin{figure}[!tb]
\centering
  \begin{subfigure}[b]{0.49\textwidth}
    \includegraphics[width=\textwidth]{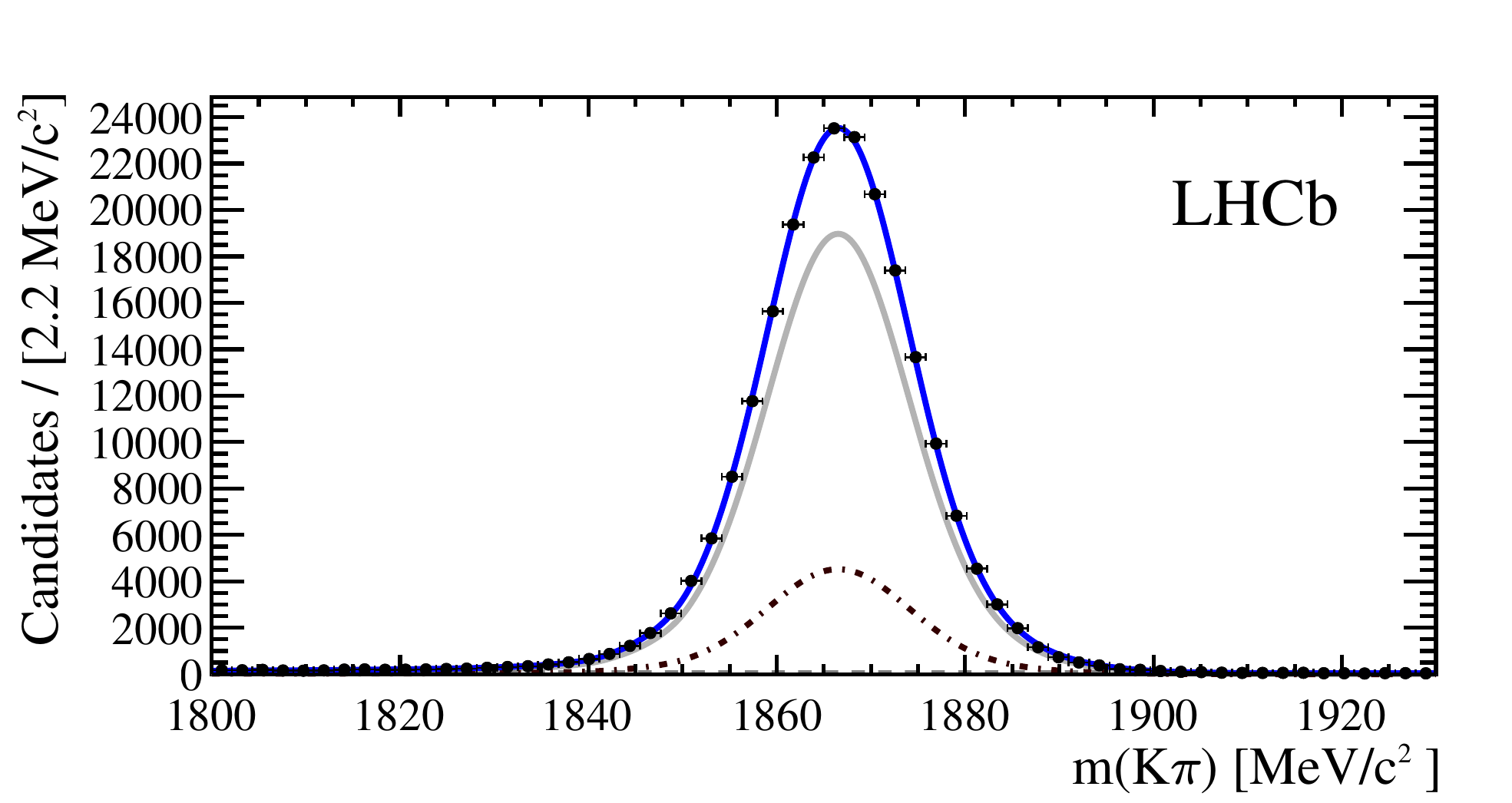}
  \end{subfigure}  
  \begin{subfigure}[b]{0.49\textwidth}
    \includegraphics[width=\textwidth]{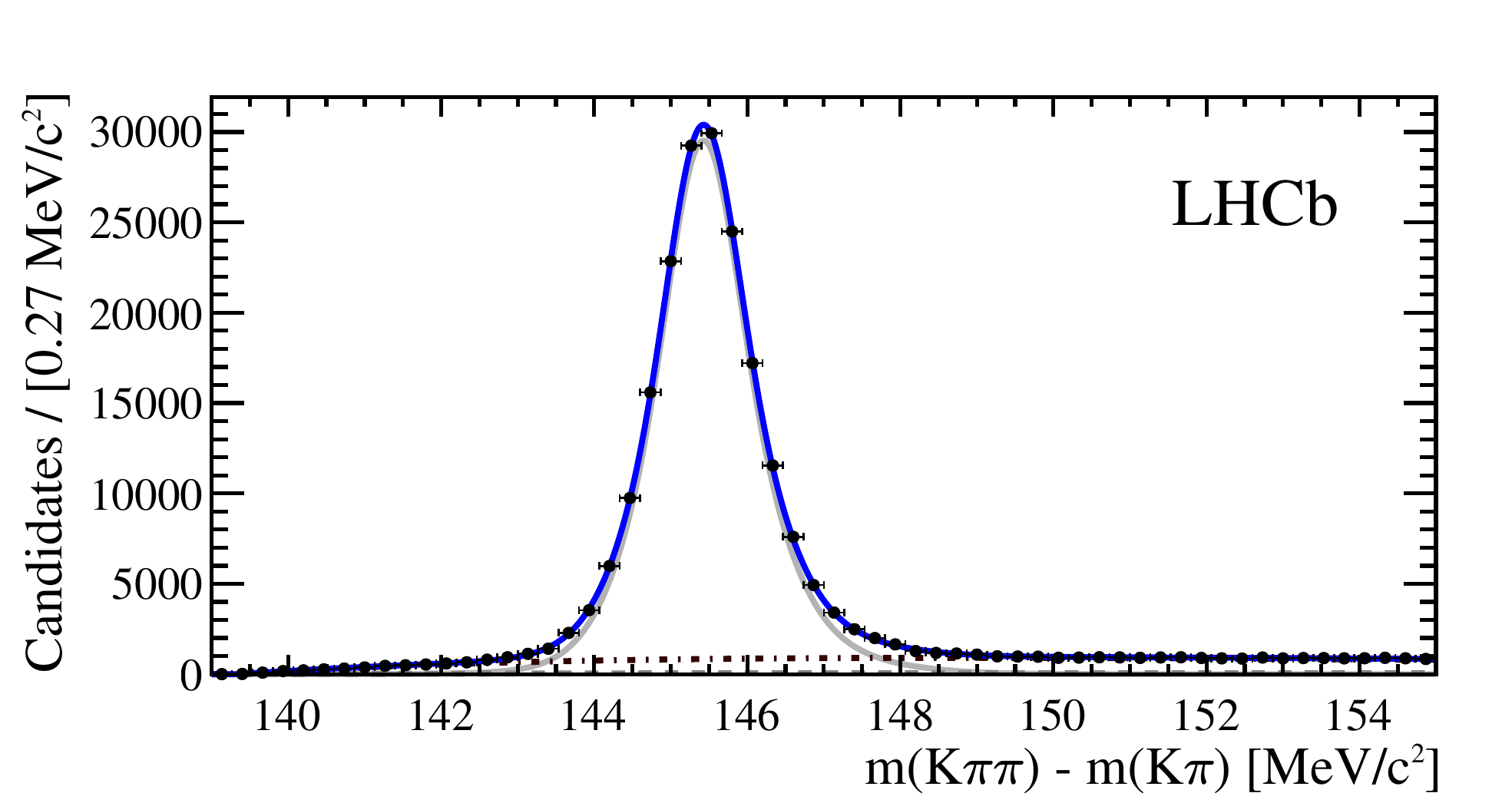}
\end{subfigure}
\caption{Distributions of (left) $m(\protect\Dz)$ and (right) $\Delta m$ for $K^-\pi^+$ candidates   
for the $8$\,\protect\tev data. The dark (blue) line shows the overall fit, the lighter grey line shows the signal, and the dot-dash
line shows genuine \protect\Dz events where the soft pion does not come from a \protect\Dstarp decay. The combinatorial
background is too small to be visible.}
\label{fig:Kpi}
\end{figure}

For the normalisation channel \DtoKpi, for which there are many candidates, binned fits are done separately to the $7$\,\tev and 
$8$\,\tev samples, using a sum of two Gaussian functions with a common mean to model the \Dz candidate mass distribution, and
a sum of three Gaussian functions for the $\Delta m$ distribution. In the latter case, two of the Gaussian functions have the
same mean, but the one with the largest width is allowed to have a different mean. The function defined by Eq.~(\ref{eq:background})
is used for the background in the $\Delta m$ spectrum, with all parameters allowed to vary in the fit. Figure~\ref{fig:Kpi} shows the results of the fit for the \DtoKpi normalization samples in 
the $8$\,\tev data, for both the $m(\Dz)$ and $\Delta m$ distributions. Totals of $80 \times 10^3$ and $182 \times 10^3$ 
{\ensuremath{\Dstarp \to \Dz (\to \Km \pip) \pip}\xspace}
decays are observed in the $7$\,\tev and $8$\,\tev data, respectively.

\section{Systematic uncertainties}
\label{sec:systematics}

The uncertainty on the fitted \Dtoemu signal rate is dominated by statistical fluctuations
of the combinatorial background. 
Sources of systematic uncertainty that could affect the final result 
include those on the yield of the normalization \DtoKpi
decay, uncertainties in the shapes of the PDFs used for \Dtoemu and \Dtopipi, 
and uncertainties in the selection efficiencies and particle misidentification probabilities.
All of these uncertainties are included as Gaussian constraints in the fits described in
Sect.~\ref{sec:Fits}.

In the nominal fit to signal candidates, the parameters of the signal PDF, obtained from 
the simulation, are Gaussian constrained according to their uncertainties. 
To obtain these uncertainties, samples of $\Bp \to \jpsi \Kp$ decays with $\jpsi \to \ep \en$ are selected 
in both simulation and data, and the $\ep \en$ mass spectra are fitted using the same functional form as 
used for \Dtoemu. The fractional differences in the parameter values between the $\jpsi \to \ep \en$ fits 
to the data and to the simulation are taken as the fractional systematic uncertainties on the corresponding 
parameters of the PDF for the \Dtoemu candidate mass spectra.

For the fits to the fully simulated, misidentified \Dtopipi mass spectra, some selection requirements are removed in
order to have enough events to obtain reliable fits. The efficiency of the selection requirements that are not 
applied varies linearly by a relative $9.4\%$ with reconstructed mass across the fit region. The PDF for the peak shape in the misidentified \Dtopipi
decays is corrected for this variation of efficiency, and the resulting contribution to the systematic uncertainty  
on the yield is taken as $4.7\%$.  

To allow for uncertainties in the fractions of \Dtoemu signal and misidentified \Dtopipi decays that are estimated 
in the three bins of BDT output,
a comparison is made between these fractions for simulated \Dtoemu, simulated \Dtopipi and well identified 
\Dtopipi decays in the data. Since the BDT does not take into account particle identification, the largest differences between
these fractions in each bin, typically $2.5\%$, are taken as the systematic uncertainties on the fractions in the data.

To account for differences between data and simulation in the per-event track multiplicity, the reconstruction efficiencies
and misidentification probabilities for simulated events are evaluated in three bins of $N_{\rm SPD}$.
These are then weighted to match the multiplicity distribution in the data. Half of the differences between the
unweighted and the weighted efficiencies and misidentification probabilities, typically $5\%$, are taken as the systematic
uncertainties on these quantities. Further uncertainties, of $2.5\%$ for each of \Dtoemu and \Dtopipi, are included 
to account for limited knowledge of the tracking efficiencies.  

Using the calibration samples, particle identification and trigger efficiencies are estimated in bins of pseudorapidity, 
transverse momentum and event multiplicity. Overall efficiencies are determined by
scaling the simulation so that the distributions in these variables match the data. To estimate systematic
uncertainties from this procedure, different binning schemes are used and the resulting changes in the 
efficiency values are treated as systematic uncertainties.
Overall systematic uncertainties are $6\%$ on the \Dtoemu selection efficiency and $30\%$ on the \Dtopipi misidentification
probability. 

To study systematic effects in the fit to the normalization channel, the order of the background polynomial is increased,
the number of bins changed, fixed parameters are varied and the Gaussian mean values in the $\Delta m$ fits
are constrained to be equal. From these studies a contribution of $1\%$ is assigned to the systematic uncertainty
on the yield. Similar procedures as described above for the signal channel are also used to evaluate the other systematic uncertainties 
for the \DtoKpi normalization channel. The resulting overall systematic uncertainty in the measured number of \DtoKpi decays is $5\%$.

\section{Results and conclusions}
\label{sec:results}

The measured branching fraction for the signal channel is given by
\begin{equation}
\BRof \Dtoemu = { \frac{N_{e\mu} / \epsilon_{e \mu}}  {{N_{K \pi} / \epsilon_{K \pi}}} } \times \BRof \DtoKpi,
\label{eq:bf}
\end{equation}
where $N_{e\mu}$ and $N_{K \pi}$ are the fitted numbers of \Dtoemu and \DtoKpi decays,
the corresponding $\epsilon$ are the overall efficiencies, and the branching fraction for the 
normalization channel, ${\BRof \DtoKpi} = (3.88 \pm 0.05)\%$, is taken from Ref.~\cite{PDG2014}.
The efficiencies $\epsilon_{e\mu} = (4.4 \pm 0.3) \times 10^{-4}$ and $\epsilon_{K\pi} = (2.5 \pm 0.1) \times 10^{-6}$, for the signal and
normalization channels, are
the products of the reconstruction efficiencies for the final-state particles,
including the geometric detector acceptance, the selection efficiencies, and the trigger efficiencies
(including the $1\%$ scaling in the trigger for the \DtoKpi channel).
 
\begin{figure}[!t]
\centering
\includegraphics[width=0.7\textwidth]{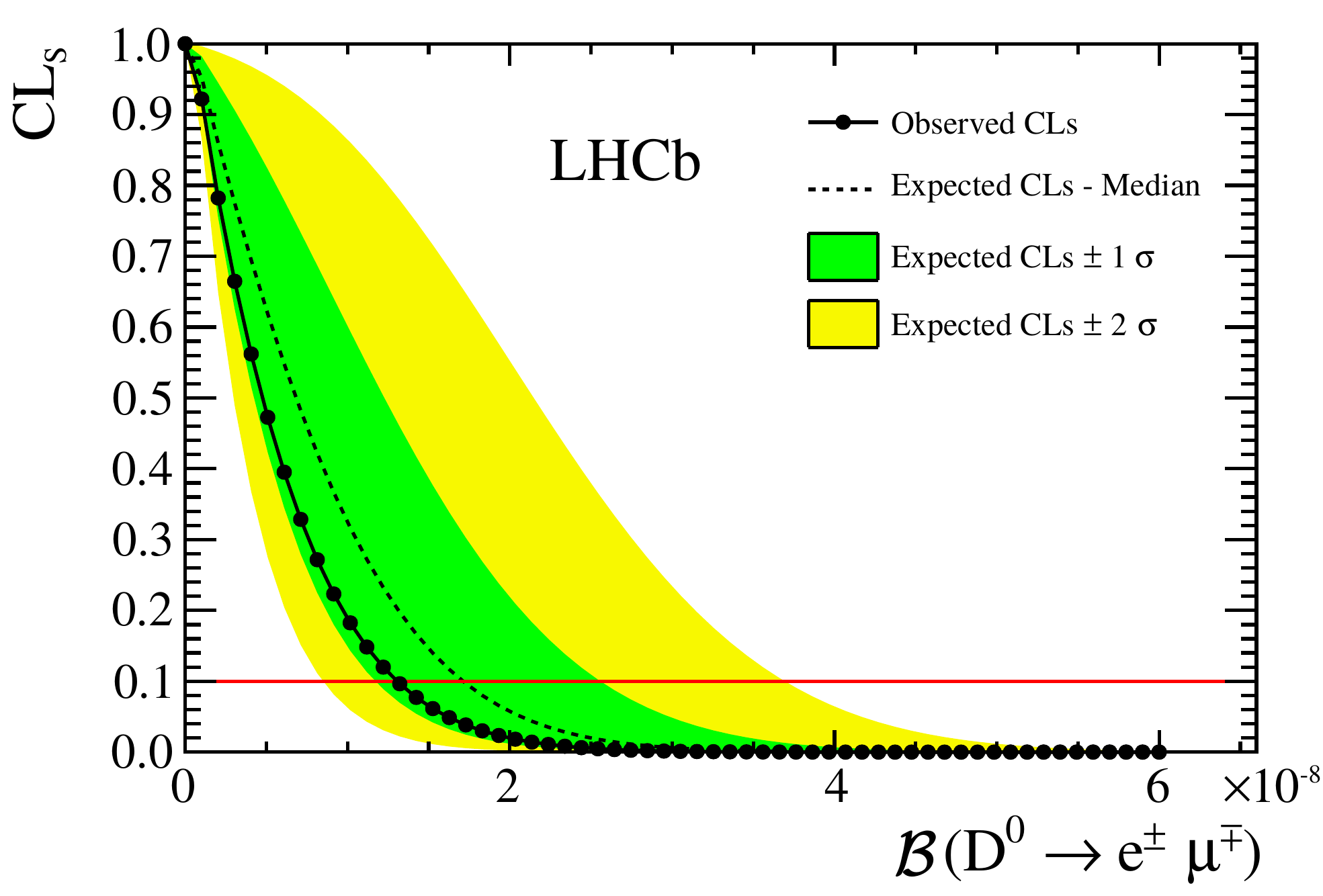}
\caption{\protect Distribution of \CLs\ as a function of \BRof \Dtoemu. The expected distribution is shown by the dashed line, 
with the $\pm 1\sigma$ and $\pm 2\sigma$
regions shaded.
The observed distribution is shown by the solid line connecting the data points. The horizontal line indicates the $90\%$ confidence level.}
\label{fig:brazil}
\end{figure}

No evidence is seen for a \Dtoemu signal in the overall mass spectrum, nor in any individual
bin of BDT output,
and the measured
branching fraction is $\BRof {\Dtoemu} = (-0.6 \pm 1.2) \times 10^{-8}$, where the uncertainty accounts for both statistical and systematic effects. 
An upper limit
on the branching fraction is obtained 
using the \CLs\ method~\cite{Read:2002hq}, 
where the p-value for the signal-plus-background hypothesis is compared
to that for the background-only hypothesis. The expected and observed \CLs\ values 
as functions of the assumed branching fraction are shown in Fig.~\ref{fig:brazil}, where the expected \CLs\ values are obtained using an
Asimov dataset~\cite{Cowan:2010js} as described in Ref.~\cite{Moneta:2010pm}, and are the median expected
limits under the assumption of no signal. 
Expected limits based on pseudoexperiments give consistent results. 
There is excellent correspondence between the expected
and observed \CLs\ values, and an upper limit is set on the branching fraction, \mbox{\BRof \Dtoemu $< 1.3  \times 10^{-8}$} 
at $90\%$ CL (and $< 1.6 \times 10^{-8}$ at $95\%$ CL). This limit will help to further constrain
products of couplings in supersymmetric models that incorporate R-parity violation~\cite{Burdman:2001tf, 
Tahir:2014ura, Wang:2014dba} and constrains the parameter space in
some leptoquark scenarios~\cite{deBoer:2015boa}.

In summary, a search for the lepton-flavour violating decay \Dtoemu is performed on a data sample corresponding to an
integrated luminosity of 3.0\invfb collected in $pp$ collisions at centre-of-mass energies of $7$  and $8$\,TeV.
The data  are consistent with the background-only hypothesis, and 
a limit is set on the branching fraction, \mbox{\BRof \Dtoemu $< 1.3  \times 10^{-8}$} 
at $90\%$ CL, which is an order of magnitude lower than the previous limit.  
 
\section*{Acknowledgements}
 
\noindent We express our gratitude to our colleagues in the CERN
accelerator departments for the excellent performance of the LHC. We
thank the technical and administrative staff at the LHCb
institutes. We acknowledge support from CERN and from the national
agencies: CAPES, CNPq, FAPERJ and FINEP (Brazil); NSFC (China);
CNRS/IN2P3 (France); BMBF, DFG and MPG (Germany); INFN (Italy); 
FOM and NWO (The Netherlands); MNiSW and NCN (Poland); MEN/IFA (Romania); 
MinES and FANO (Russia); MinECo (Spain); SNSF and SER (Switzerland); 
NASU (Ukraine); STFC (United Kingdom); NSF (USA).
We acknowledge the computing resources that are provided by CERN, IN2P3 (France), KIT and DESY (Germany), INFN (Italy), SURF (The Netherlands), PIC (Spain), GridPP (United Kingdom), RRCKI (Russia), CSCS (Switzerland), IFIN-HH (Romania), CBPF (Brazil), PL-GRID (Poland) and OSC (USA). We are indebted to the communities behind the multiple open 
source software packages on which we depend. We are also thankful for the 
computing resources and the access to software R\&D tools provided by Yandex LLC (Russia).
Individual groups or members have received support from AvH Foundation (Germany),
EPLANET, Marie Sk\l{}odowska-Curie Actions and ERC (European Union), 
Conseil G\'{e}n\'{e}ral de Haute-Savoie, Labex ENIGMASS and OCEVU, 
R\'{e}gion Auvergne (France), RFBR (Russia), GVA, XuntaGal and GENCAT (Spain), The Royal Society 
and Royal Commission for the Exhibition of 1851 (United Kingdom).

\addcontentsline{toc}{section}{References}
\setboolean{inbibliography}{true}
\bibliographystyle{LHCb}
\bibliography{main,LHCb-PAPER,LHCb-CONF,LHCb-DP,LHCb-TDR,Our}

\newpage


\centerline{\large\bf LHCb collaboration}
\begin{flushleft}
\small
R.~Aaij$^{39}$, 
C.~Abell\'{a}n~Beteta$^{41}$, 
B.~Adeva$^{38}$, 
M.~Adinolfi$^{47}$, 
A.~Affolder$^{53}$, 
Z.~Ajaltouni$^{5}$, 
S.~Akar$^{6}$, 
J.~Albrecht$^{10}$, 
F.~Alessio$^{39}$, 
M.~Alexander$^{52}$
% ...

{\footnotesize \it
% ...
$ ^{5}$Clermont Universit\'{e}, Universit\'{e} Blaise Pascal, CNRS/IN2P3, LPC, Clermont-Ferrand, France\\
$ ^{6}$CPPM, Aix-Marseille Universit\'{e}, CNRS/IN2P3, Marseille, France\\
% ...
$ ^{10}$Fakult\"{a}t Physik, Technische Universit\"{a}t Dortmund, Dortmund, Germany\\
% ...
$ ^{38}$Universidad de Santiago de Compostela, Santiago de Compostela, Spain\\
$ ^{39}$European Organization for Nuclear Research (CERN), Geneva, Switzerland\\
% ...
$ ^{41}$Physik-Institut, Universit\"{a}t Z\"{u}rich, Z\"{u}rich, Switzerland\\
% ...
$ ^{47}$H.H. Wills Physics Laboratory, University of Bristol, Bristol, United Kingdom\\
% ...
$ ^{52}$School of Physics and Astronomy, University of Glasgow, Glasgow, United Kingdom\\
$ ^{53}$Oliver Lodge Laboratory, University of Liverpool, Liverpool, United Kingdom\\
% ...
}
\end{flushleft}

\end{document}

%% file: lhcb-symbols-def.tex
\def\lhcb {\mbox{LHCb}\xspace}

\def\MagUp {\mbox{\em Mag\kern -0.05em Up}\xspace}

\ifthenelse{\boolean{uprightparticles}}%
{

 \def\Pmu         {\ensuremath{\upmu}\xspace}                 
 \def\Pnu         {\ensuremath{\upnu}\xspace}                 
                  
 \def\Ppi         {\ensuremath{\uppi}\xspace}

 \def\Ppsi        {\ensuremath{\uppsi}\xspace}

 \def\PDelta      {\ensuremath{\Delta}\xspace}                 
 \def\PXi      {\ensuremath{\Xi}\xspace}                 
 \def\PLambda      {\ensuremath{\Lambda}\xspace}                 
 \def\PSigma      {\ensuremath{\Sigma}\xspace}                 
 \def\POmega      {\ensuremath{\Omega}\xspace}                 
 \def\PUpsilon      {\ensuremath{\Upsilon}\xspace}                 
 
 \def\PB      {\ensuremath{\mathrm{B}}\xspace}                 
                  
 \def\PD      {\ensuremath{\mathrm{D}}\xspace}

 \def\PJ      {\ensuremath{\mathrm{J}}\xspace}                 
 \def\PK      {\ensuremath{\mathrm{K}}\xspace}

 \def\Pb      {\ensuremath{\mathrm{b}}\xspace}                 
 \def\Pc      {\ensuremath{\mathrm{c}}\xspace}                 
                  
 \def\Pe      {\ensuremath{\mathrm{e}}\xspace}

 \def\Pi      {\ensuremath{\mathrm{i}}\xspace}

 \def\Ps      {\ensuremath{\mathrm{s}}\xspace}

}
{

 \def\Pmu         {\ensuremath{\mu}\xspace}                 
 \def\Pnu         {\ensuremath{\nu}\xspace}                 
                  
 \def\Ppi         {\ensuremath{\pi}\xspace}

 \def\Ppsi        {\ensuremath{\psi}\xspace}                 
                  
 \mathchardef\PDelta="7101
 \mathchardef\PXi="7104
 \mathchardef\PLambda="7103
 \mathchardef\PSigma="7106
 \mathchardef\POmega="710A
 \mathchardef\PUpsilon="7107
                  
 \def\PB      {\ensuremath{B}\xspace}                 
                  
 \def\PD      {\ensuremath{D}\xspace}

 \def\PJ      {\ensuremath{J}\xspace}                 
 \def\PK      {\ensuremath{K}\xspace}

 \def\Pb      {\ensuremath{b}\xspace}                 
 \def\Pc      {\ensuremath{c}\xspace}                 
                  
 \def\Pe      {\ensuremath{e}\xspace}

 \def\Pi      {\ensuremath{i}\xspace}

 \def\Ps      {\ensuremath{s}\xspace}

}

\makeatletter
\ifcase \@ptsize \relax
  \newcommand{\miniscule}{\@setfontsize\miniscule{4}{5}}
\or
  \newcommand{\miniscule}{\@setfontsize\miniscule{5}{6}}
\or
  \newcommand{\miniscule}{\@setfontsize\miniscule{5}{6}}
\fi
\makeatother

\DeclareRobustCommand{\optbar}[1]{\shortstack{{\miniscule (\rule[.5ex]{1.25em}{.18mm})}
  \\ [-.7ex] $#1$}}

\def\en         {{\ensuremath{\Pe^-}}\xspace}   
\def\ep         {{\ensuremath{\Pe^+}}\xspace}
\def\epm        {{\ensuremath{\Pe^\pm}}\xspace}

\def\mup        {{\ensuremath{\Pmu^+}}\xspace}

\def\neu        {{\ensuremath{\Pnu}}\xspace}

\def\neue       {{\ensuremath{\neu_e}}\xspace}

\def\neum       {{\ensuremath{\neu_\mu}}\xspace}

\def\squark    {{\ensuremath{\Ps}}\xspace}

\def\cquark    {{\ensuremath{\Pc}}\xspace}

\def\bquark    {{\ensuremath{\Pb}}\xspace}

\def\pion   {{\ensuremath{\Ppi}}\xspace}

\def\pip    {{\ensuremath{\pion^+}}\xspace}
\def\pim    {{\ensuremath{\pion^-}}\xspace}

\def\kaon    {{\ensuremath{\PK}}\xspace}
  \def\Kbar    {{\kern 0.2em\overline{\kern -0.2em \PK}{}}\xspace}

\def\KorKbar    {\kern 0.18em\optbar{\kern -0.18em K}{}\xspace}

\def\Kp      {{\ensuremath{\kaon^+}}\xspace}
\def\Km      {{\ensuremath{\kaon^-}}\xspace}

  \def\Dbar    {{\kern 0.2em\overline{\kern -0.2em \PD}{}}\xspace}
\def\D       {{\ensuremath{\PD}}\xspace}

\def\DorDbar    {\kern 0.18em\optbar{\kern -0.18em D}{}\xspace}
\def\Dz      {{\ensuremath{\D^0}}\xspace}

\def\Dp      {{\ensuremath{\D^+}}\xspace}

\def\Dstarp  {{\ensuremath{\D^{*+}}}\xspace}

\def\Dsp     {{\ensuremath{\D^+_\squark}}\xspace}

\def\B       {{\ensuremath{\PB}}\xspace}
\def\Bbar    {{\ensuremath{\kern 0.18em\overline{\kern -0.18em \PB}{}}}\xspace}

\def\BorBbar    {\kern 0.18em\optbar{\kern -0.18em B}{}\xspace}

\def\Bu      {{\ensuremath{\B^+}}\xspace}

\def\Bp      {{\ensuremath{\Bu}}\xspace}

\def\jpsi     {{\ensuremath{{\PJ\mskip -3mu/\mskip -2mu\Ppsi\mskip 2mu}}}\xspace}

  \def\Y#1S{\ensuremath{\PUpsilon{(#1S)}}\xspace}

\def\Lbar        {{\ensuremath{\kern 0.1em\overline{\kern -0.1em\PLambda}}}\xspace}
\def\LorLbar    {\kern 0.18em\optbar{\kern -0.18em \PLambda}{}\xspace}

\def\to                 {\ensuremath{\rightarrow}\xspace}

\def\AT#1     {\ensuremath{A_{\mathrm{T}}^{#1}}\xspace}           

\def\C#1      {\ensuremath{\mathcal{C}_{#1}}\xspace}                       
\def\Cp#1     {\ensuremath{\mathcal{C}_{#1}^{'}}\xspace}                    
\def\Ceff#1   {\ensuremath{\mathcal{C}_{#1}^{\mathrm{(eff)}}}\xspace}        
\def\Cpeff#1  {\ensuremath{\mathcal{C}_{#1}^{'\mathrm{(eff)}}}\xspace}       
\def\Ope#1    {\ensuremath{\mathcal{O}_{#1}}\xspace}                       
\def\Opep#1   {\ensuremath{\mathcal{O}_{#1}^{'}}\xspace}                    

\newcommand{\tev}{\ifthenelse{\boolean{inbibliography}}{\ensuremath{~T\kern -0.05em eV}\xspace}{\ensuremath{\mathrm{\,Te\kern -0.1em V}}}\xspace}
\newcommand{\gev}{\ensuremath{\mathrm{\,Ge\kern -0.1em V}}\xspace}
\newcommand{\mev}{\ensuremath{\mathrm{\,Me\kern -0.1em V}}\xspace}
\newcommand{\kev}{\ensuremath{\mathrm{\,ke\kern -0.1em V}}\xspace}
\newcommand{\ev}{\ensuremath{\mathrm{\,e\kern -0.1em V}}\xspace}
\newcommand{\gevc}{\ensuremath{{\mathrm{\,Ge\kern -0.1em V\!/}c}}\xspace}
\newcommand{\mevc}{\ensuremath{{\mathrm{\,Me\kern -0.1em V\!/}c}}\xspace}
\newcommand{\gevcc}{\ensuremath{{\mathrm{\,Ge\kern -0.1em V\!/}c^2}}\xspace}
\newcommand{\gevgevcccc}{\ensuremath{{\mathrm{\,Ge\kern -0.1em V^2\!/}c^4}}\xspace}
\newcommand{\mevcc}{\ensuremath{{\mathrm{\,Me\kern -0.1em V\!/}c^2}}\xspace}

\def\mum  {\ensuremath{{\,\upmu\rm m}}\xspace}

\def\invfb   {\ensuremath{\mbox{\,fb}^{-1}}\xspace}

\newcommand{\chisqip}{\ensuremath{\chi^2_{\rm IP}}\xspace}

\def\gsim{{~\raise.15em\hbox{$>$}\kern-.85em
          \lower.35em\hbox{$\sim$}~}\xspace}
\def\lsim{{~\raise.15em\hbox{$<$}\kern-.85em
          \lower.35em\hbox{$\sim$}~}\xspace}

\def\ptot       {\mbox{$p$}\xspace}
\def\pt         {\mbox{$p_{\rm T}$}\xspace}

\def\evtgen     {\mbox{\textsc{EvtGen}}\xspace}

\def\geant      {\mbox{\textsc{Geant4}}\xspace}

\def\photos     {\mbox{\textsc{Photos}}\xspace}

\def\pythia     {\mbox{\textsc{Pythia}}\xspace}

\def\tell1  {TELL1\xspace}
\def\ukl1   {UKL1\xspace}

